\documentclass{nature_withfigs}

\usepackage{graphicx}
\usepackage{verbatim}
\usepackage{hhline}
\usepackage{amsfonts,amssymb,amsmath,amscd}
\usepackage{graphics,epsfig,graphicx}
\usepackage{amsfonts,amssymb,latexsym,amsmath,amscd}
\usepackage{times,mathptm,color}
\usepackage{color}
\usepackage{latexsym}

\title{Evidence for a Bose-Einstein condensate of excitons}

\author{Mathieu Alloing$^{1}$, Mussie Beian$^{1}$, Maciej Lewenstein$^{1,2}$, David Fuster$^{3}$, Yolanda
Gonz\' {a}lez$^{3}$, Luisa Gonz\' {a}lez$^{3}$,Roland Combescot$^{4,5}$,
Monique Combescot$^{6}$ and Fran\c{c}ois Dubin$^{1,6}$}

\begin{document}

\maketitle

\begin{affiliations}

\item  ICFO-The Institute of Photonic Sciences,
Av. Carl Friedrich Gauss 3, 08860 Castelldefels, Spain
\item ICREA-Instituci\'{o} Catalana de Recerca i Estudis Avanats, 
Lluis Companys 23, 08010 Barcelona, Spain
\item IMM-Instituto de Microelectr\'{o}nica de Madrid (CNM-CSIC), 
Isaac Newton 8, PTM, E-28760 Tres Cantos, Madrid, Spain
\item  Laboratoire de Physique Statistique, Ecole Normale Sup\' {e}rieure, 
UPMC Paris 06, Universit\' {e} Paris Diderot, CNRS, 24 rue Lhomond, 75005 Paris,
France
\item  Institut Universitaire de France, 103 boulevard Saint-Michel, 75005
Paris,
France
\item Institut des Nanosciences de Paris, CNRS and UPMC, 2 pl. Jussieu,
75005 Paris, France
\end{affiliations}


\begin{abstract}

The demonstration of Bose-Einstein condensation in atomic gases at
micro-Kelvin temperatures is a striking landmark\cite{Stringari_Book_BEC} while
its evidence for semiconductor
excitons\cite{Moskalenko_62,Blatt_1962,Keldysh_BCS,Keldysh_BEC} still is a
long-awaited milestone.
This situation was not foreseen because excitons are light-mass boson-like
particles with a condensation expected to occur around a few Kelvins\cite{Stringari_Book,Snoke_Book}. 
An explanation can be found in the underlying fermionic nature of excitons which rules their
condensation\cite{Monique_dark_BEC}. Precisely, it was recently predicted that, at accessible experimental conditions, the exciton condensate shall be "gray" with a dominant dark part 
coherently coupled to a weak bright component through fermion exchanges\cite{Roland_2012}. This counter-intuitive quantum condensation, since excitons are mostly known for their optical activity, directly follows from the excitons internal structure which has an optically inactive, i.e., dark, ground state\cite{Monique_dark_BEC}. Here, we report compelling evidence for such
a "gray" condensate. We use an all-optical approach in order to
produce microscopic traps which confine a dense exciton gas that yet exhibits an
anomalously weak photo-emission at sub-Kelvin temperatures. This first fingerprint for 
a "gray" condensate is then confirmed by the macroscopic spatial
coherence and the linear polarization of the weak excitonic photoluminescence
emitted from the trap, as theoretically predicted\cite{Roland_2012}. 

\end{abstract}


Bose-Einstein condensation of semiconductor
excitons has received a considerable attention since 
its theoretical prediction
\cite{Moskalenko_62,Blatt_1962,Keldysh_BCS,Keldysh_BEC} in the 1960's.
Unlike most commonly studied bosonic atoms\cite{Stringari_Book},
the exciton composite nature plays a key role in their condensation.
In semiconductor quantum wells, excitons are made of spin ($\pm$1/2) electrons 
and ''spin'' ($\pm$3/2) holes. These carriers
mainly interact through attractive intraband Coulomb processes. Weak interband
valence-conduction Coulomb processes also
exist, but only for optically active, or "bright",
excitons with total spin
($\pm$1) made of ($\pm$3/2)
holes and ($\mp$1/2) electrons. These (repulsive) interband processes bring the
energy of
bright excitons above the one of ``dark`` excitons with total spin
($\pm$2) made of ($\pm$3/2) holes and ($\pm$1/2) electrons. As a result,
Bose-Einstein condensation of excitons must occur within the low energy dark
states  \cite{Monique_dark_BEC,Monique_Leuenberger}.
This feature, which makes exciton condensation hard
to evidence by conventional optical means, could be the reason 
why unambiguous signatures of exciton condensates have not been
given yet, despite several decades of active experimental research. By contrast,
the
excitonic component of a
polariton being by construction coupled to light, polariton condensation 
can be studied through photoluminescence, which recently led to
remarkable experiments\cite{Deng_2002,Kasprzak_2006,Wertz_2010,Balili_2007}.

While in the very dilute regime the exciton condensate must be completely dark, no
matter how small the energy difference between dark and
bright states is \cite{Monique_dark_BEC}, it was recently
shown\cite{Roland_2012} that at sufficiently large density,
carrier exchange between bright and dark excitons brings a coherent
bright component to the condensate which becomes "gray". 
Such a coherence between dark and bright excitons is very similar to what
occurs in the well known
phases of superfluid $^3$He (see e.g. Ref.\cite{Leggett_2006}), and in the more
recent spinor condensates of ultracold
atomic Bose gases\cite{Ueda_2013}, where components of these superfluids
corresponding to different internal degrees
of freedom coexist and are coherent. This coherent coupling allows probing the
exciton
condensate through the photoluminescence of its
bright part. As the bright component is very small, the photoluminescence signal
is very weak. Nevertheless, it shall unveil the existence of a dense population of
dark excitons, 
the spatial coherence of the condensate and its internal "spin"
structure through the polarisation of the emitted light.
In semiconductor quantum wells, the
dark nature of exciton Bose-Einstein condensation has been overlooked
until very recently\cite{Monique_Leuenberger,Monique_dark_BEC}, most probably because
the splitting between bright and dark states is small compared to the thermal
energy at critical temperature. Here, we present compelling experimental evidence 
for a "gray" Bose-Einstein condensate of excitons\cite{Monique_dark_BEC,Roland_2012} through the experimental observation of all its theoretically predicted characteristics.  

We study excitons confined in a 25 nm wide GaAs single quantum well embedded in
a field-effect device with an electrical polarization keeping
electrons and holes well apart. As the electron and hole wave functions have a
small
overlap, these dipolar excitons have a long radiative lifetime ($\sim$ 20
ns) and a rather large energy splitting between bright and dark
states ($\sim$ 20 $\mu$eV, see Ref. \cite{Blackwood_94,Timofeev_2013}). The
electrical polarisation also ensures a repulsive effective exciton-exciton
interaction which prevents the formation of biexcitons at the typical density $n_c$
where Bose statistics becomes dominant ($n_c \simeq m_X k_B T/\hbar^2
\sim$ 10$^{10}$ cm$^{-2}$  for 2D excitons with mass $m_X$ at 1K).

We use a pump laser ($\lambda_\mathrm{pump}$=641.5 nm) 
with an energy above the AlGaAs barriers of the quantum
well  in order to create a dense and well
thermalised exciton gas. For such laser excitation photo-injected electrons and holes are captured by the
quantum well with different efficiencies\cite{Butov_2004,Rapaport_2004}: 
a region richer in holes is formed
around the laser excitation, itself surrounded by an electron-rich domain
resulting from both the photo-current passing through our device and the
modulation doping of the structure (see Fig. 1.A). In this
landscape, dipolar excitons are created through the Coulomb interaction
between photo-injected electrons and holes, the exciton transport being somewhat complicated 
by the ambipolar diffusion of excess carriers which screen the external field applied through our top gate electrode.

Figures 1.C and 1.D show the photoluminescence emission 10 ns after the pump
pulse at 350 mK and 7K respectively. Both reveal a pattern characteristic of the charge separation existing in the quantum well, namely a macroscopic exciton ring formed a few tens of microns away from the pump excitation\cite{Butov_2002,Snoke_2002}. Figures 1.E and 1.F show the exciton confinement potential together with the profile of the exciton density for these measurements. They are both deduced from a weak probe pulse which injects a very dilute exciton cloud after the pump pulse. Note that this probe beam is tuned well below the bandgap of the AlGaAs barriers ($\lambda_\mathrm{probe}$=790 nm) in order to bring essentially no perturbation\cite{Alloing_Sci_rep}. Excitons injected by the probe beam emit a photoluminescence at an energy $E_X\simeq$($\overrightarrow{d}.\overrightarrow{F}_\mathrm{screen}+u_0n_X$). The first term corresponds to the energy increase of a single exciton resulting from the screening field \textbf{F}$_\mathrm{screen}$ induced by excess carriers, $d\approx e.15$~nm being the excitonic dipole moment\cite{Supplements}. The 
second term, where $n_X$ is the exciton density and $u_0$ a parameter given by the
geometry of our heterostructure \cite{Ivanov_2010}, corresponds to the energy increase resulting from repulsive exciton-exciton interactions. By sending the probe pulse 100 ns after extinction of the pump, i.e. when all excitons injected by the pump have recombined, we directly map the excitons confinement as for very small exciton densities $E_X$ reduces to $E_X^{\mathrm{(probe)}}\,=\,\overrightarrow{d}.\overrightarrow{F}_\mathrm{screen}$. We can estimate the density profile of excitons produced by the pump excitation by sending a probe pulse 10 ns after extinction of the pump as $E_X$ reduces then to   $E_X^{\mathrm{(pump)}}$=($E_X^{\mathrm{(probe)}}$+$u_0$n$_X$). This pump-probe spectroscopy then allows us to extract the whole exciton density injected by the pump pulse, n$_X$, including dark and bright excitons, a crucial ingredient to signal a "gray" condensate.

Figures 1.E-F show the profile of the exciton confinement through $E_X^{\mathrm{(probe)}}$. It displays a maximum at the pump laser spot, then slowly decreases with the distance to the laser excitation before an abrupt decrease just before the region where the ring is formed. This is physically reasonable because the ring is expected to be located at the interface between the electron-rich
and hole-rich domains where the near absence of excess carriers leads to a minimum screening.
Most importantly, our measurements reveal at 350 mK that an electrostatic trap, i.e. a local minimum of $E_X^{\mathrm{(probe)}}$, is 
spontaneously formed in the vicinity of the ring (see Fig. 1.E). This potential confines dipolar excitons which are high-field seekers, i.e., attracted by strong field regions. We note that the exciton trap is not homogeneous along the circumference of the ring, i.e., it is not identical on opposite sides (see Fig. 1.E). Its depth can be as large as $\simeq$ 1.2 meV at 350 mK, an increase of the bath temperature leading to a reduction or a total suppression of the trap (Fig. 1.F).  

We evaluate the exciton density by comparing $E_X^{\mathrm{(pump)}}$ to $E_X^{\mathrm{(probe)}}$. Figure 1.E shows at 350 mK that  across the emission $E_X^{\mathrm{(pump)}}$ is
blueshifted compared to $E_X^{\mathrm{(probe)}}$. This reveals that the pump pulse creates a dense  exciton gas (n$_X$$\sim$ 10$^{10}$ cm$^{-2}$ for a blueshift ($E_X^{\mathrm{(pump)}}$-$E_X^{\mathrm{(probe)}}$) $\sim$ 1meV). More strikingly, we also note that $E_X^{\mathrm{(pump)}}$ is almost constant in the region of the ring (Fig. 1.E), unlike at 7K (Fig. 1.F). This reveals that dipolar excitons  completely fill the trap spontaneously formed in the vicinity of the ring. The experimental blueshift (1.2 meV around the position -25 $\mu$m in Fig. 1.E) leads to an exciton density $\sim$ 2 10$^{10}$ cm$^{-2}$ across the entire trap \cite{Ivanov_2010,Schindler_08,Rapaport_09}. At the same time, the photoluminescence intensity is reduced by more than ten fold between the positions (-25) and (-31) $\mu$m. At the latter position, the excitonic population then dominantly consists of optically inactive, i.e. dark, excitons. The dense but nearly dark exciton gas, highlighted by the gray area in Fig. 1.E,  can only be explained by the formation of a condensate which acts as a sink and captures most of the excitons, this condensate being nearly dark, or "gray". Indeed, in the absence of a condensate, i.e. in the classical regime, the dark and bright populations should be very similar, because the dark-bright energy
splitting ($\sim$20 $\mu$eV) is smaller than the thermal energy ($ \sim$ 30 $\mu$eV at our
lowest bath temperature). This would lead to a much stronger photoluminescence than the one we observe.
Such a conclusion is further supported by the spectral
profile of the photoluminescence which is essentially identical on the brightest
part of the ring and 6 $\mu$m outside\cite{Supplements}, fully coherent with the fact that the total exciton density stays unchanged throughout this region.
Finally, Fig. 1-F shows that at higher temperatures we no longer find the spectral signature of a "gray"
condensate. We wish note that the specific position where the condensate is
formed certainly is the result of a complex hydrodynamical diffusion, 
dipolar excitons experiencing a chute from the pump laser spot to the trapping region,
while cooling at the same time.

To unambiguously confirm the formation of a "gray" condensate at sub-Kelvin bath temperatures, we further study the first order spatial coherence $|g^{(1)}|$ of the
photoluminescence\cite{Supplements} since the light emitted by the 
bright part of the condensate reflects its long-range order. 
The classical regime is distinguished from the quantum regime
through its spatial coherence\cite{Glauber_1999,IB_2000}, the classical coherence
length being of the order of the de Broglie wavelength
($\lambda_\mathrm{dB}\sim$ 100 nm
at sub-Kelvin temperatures). We experimentally assess 
the coherence length of bright excitons by using an actively stabilized
Mach-Zehnder interferometer. One arm of the
interferometer displaces the photoluminescence laterally by $\delta_x$=1.5
$\mu$m with respect to the second
arm; it also tilts it vertically, so that output interference fringes end up
aligned horizontally\cite{Supplements,High_2012}. By scanning the phase of the
interferometer, we reconstruct point by point the amplitude of the
interference contrast. This allows us to draw the map of the emission first order
spatial coherence from which we deduce\cite{Supplements} the exciton coherence
length $\xi$.

Figure 2 shows $|g^{(1)}|$ in
the ring region. At high temperature ($T_\mathrm{b}$= 7K in Fig. 2.D),
the interference contrast does not significantly vary across the emission,
staying approximately equal to
10-15 $\%$ which is the background value of our interference contrast
\cite{Supplements}. 
This leads to a coherence length $\xi\lesssim$ 200
nm. At lower temperature ($T_\mathrm{b}$= 370 mK in Fig. 2.C), the interference
contrast exhibits a pattern correlated with the spatial
profile of the photoemission in a way which again reveals a 
"gray" condensate. Indeed, Fig. 3.A shows that
the interference contrast is minimal ($\sim$10 $\%$) in the brightest part of
the ring while, in the outer region where the photoluminescence intensity is strongly decreased, 
$|g^{(1)}|$  can reach $\approx$ 40 $\%$ which is above half the
auto-correlation  value
(70$\%$ for $\delta_x=0$). This shows that the observed bright excitons with 
a coherence length $\xi\sim$ 1.5 $\mu$m 
one order of magnitude larger than the de Broglie wavelength, belong to the condensate. The variation of
the excitonic coherence as a function of the
bath temperature moreover shows that non-classical correlations become dominant 
below a critical temperature $\sim$2K, the coherence
length abruptly increasing from near classical to non-classical values (see Fig. 3.C).

 In order to obtain further insight into the internal structure of a gray
condensate,
we filter the polarization of the
photoluminescence.
In a "gray" condensate, dark and bright states are coupled by carrier
exchanges\cite{Roland_2012} or other coupling processes as suggested recently \cite{Ali_2009,Kavokin_2012,Malpuech_2013,High_2012}.
Since the lowest energy for degenerate ($\pm$1) or ($\pm$2) states is obtained for 
a linear polarisation, due again to
carrier exchanges\cite{Monique_dark_BEC}, a "gray" condensate must exhibit a linearly
polarized photoluminescence. And
indeed, Fig. 4.B shows at 370 mK that the photoluminescence is mostly polarized linearly in the
outer region of the ring where macroscopic coherence is also observed
 (Fig. 3.C) -- the degree of circular polarization being far smaller (see Fig.4.C). 
In the inner region of the ring, we also observe a 
linear polarization, but along the orthogonal direction. 
These observations contrast with recent studies performed
in double quantum well heterostructures where dipolar (spatially indirect)
excitons exhibit
correlated patterns having both linear and circular polarizations\cite{High_2013}.
These patterns have been interpreted in terms
of coherent exciton transport and spin-orbit
coupling\cite{High_2012,Kavokin_2012,Malpuech_2013}.
Our experiments, which do not reveal these patterns, are performed in a single
quantum well with a dark-bright energy splitting much larger than in
bilayer heterostructures. Since this splitting plays a key role in
selecting the specific condensate which is formed, it is reasonable to think
that
experiments performed in single and double quantum wells probe different
regimes.

As a last remark, one might wonder if our essentially
two-dimensional geometry would not dramatically affect  Bose-Einstein
condensation. This is not so
because condensation occurs in small regions which stabilizes the
condensate\cite{Stringari_Book_BEC}.
In addition, phase fluctuations, responsible for the main
qualitative differences between 2D and 3D systems\cite{Stringari_Book_BEC}, are
then quenched. Hence the condensate should be fairly similar to a 3D condensate.






\clearpage
\centerline{\scalebox{1.1}{\includegraphics{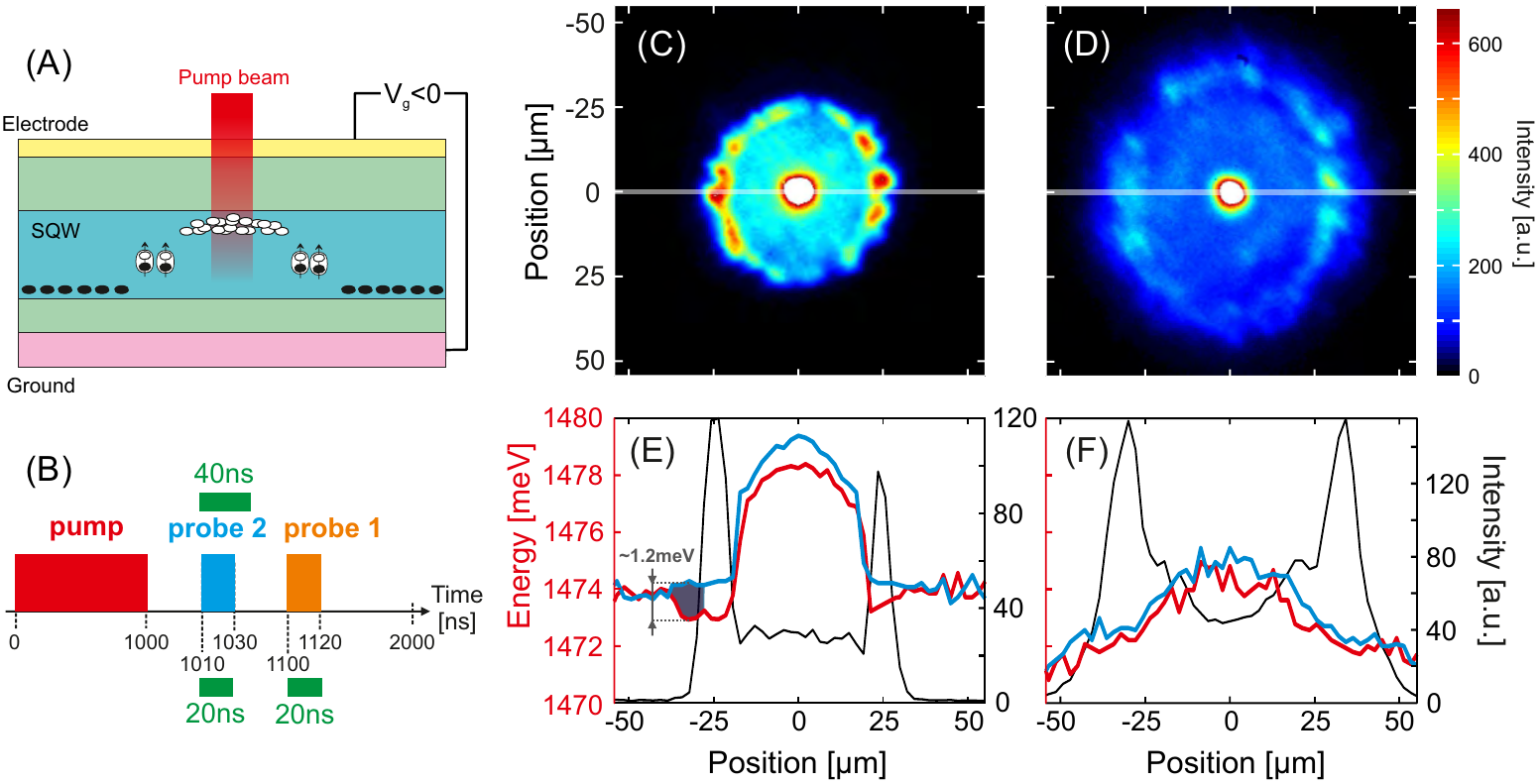}}}
\noindent {\bf Fig. 1.} (A): Sketch of our field-effect device: a pump beam excites a wide single quantum well (SQW) and produces a region rich in holes (open circles) around the laser spot, surrounded by a region rich in electrons (filled circles). These excess charges screen the electric field imposed by the potential V$_g$ applied on the top electrode of the device. (B): Schematic time sequence showing pump and probe pulses together with the intervals during which our experimental results are recorded (green). (C-D) Real photoluminescence image recorded in a 40ns long time interval, starting 10 ns after extinction of  the pump excitation, at 350 mK (C) and at 7K (D). (E-F) Spatial profiles of the excitonic confinement deduced from the 
$E_X^{\mathrm{(probe)}}$ measured in the "probe 1" pulse (red line) together with the exciton blueshift
deduced from $E_X^{\mathrm{(pump)}}$ measured during the "probe 2" pulse (blue line). These profiles are taken along the white straight lines indicated in (C-D). The solid black lines in (E-F) show the profiles of the photoluminescence intensity. In (E) the gray region underlines a large exciton density at 350 mK in a region of anomalously weak photoluminescence. It reveals the existence of a "gray" exciton condensate.

\clearpage
\begin{center}
\includegraphics[width=.9\textwidth]{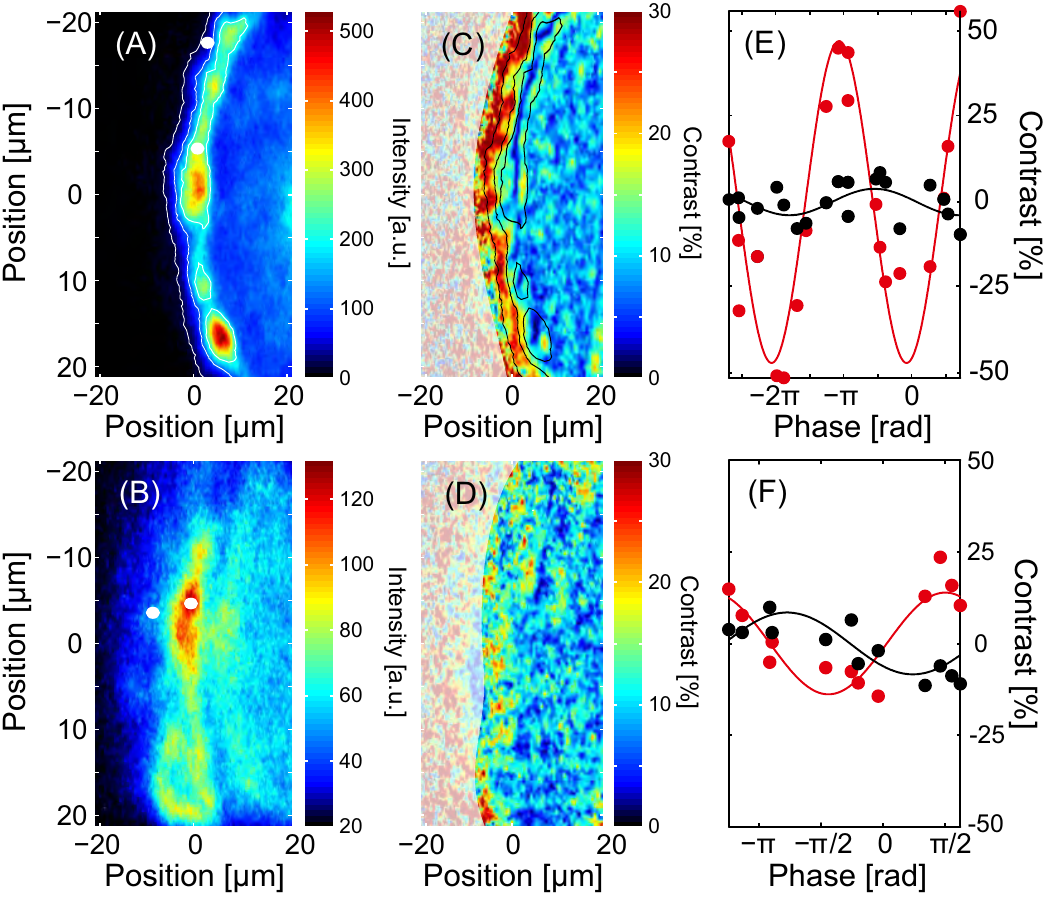}
\end{center}
\noindent {\bf Fig. 2.} (A-B): Photoluminescence taken from the same position of the
ring at 370 mK (A) and at 7K (B). 
(C-D): Map of the interference contrast for $\delta_x$=1.5 $\mu$m at 370 mK (C)
and at 7K (D). The shaded white area masks the region where the interference contrast can not be accurately measured because the photoluminescence intensity is too weak.
(E-F): Variation of the interference
contrast, at 370 mK (E) and at 7 K (F), as a function of the
interferometer phase. Data are taken at the
position of the ring (black) and 5 $\mu$m outside the ring (red). The
white circles in (A) and (B) show these two positions.

\clearpage

\centerline{\scalebox{1}{\includegraphics[width=1.1\textwidth]{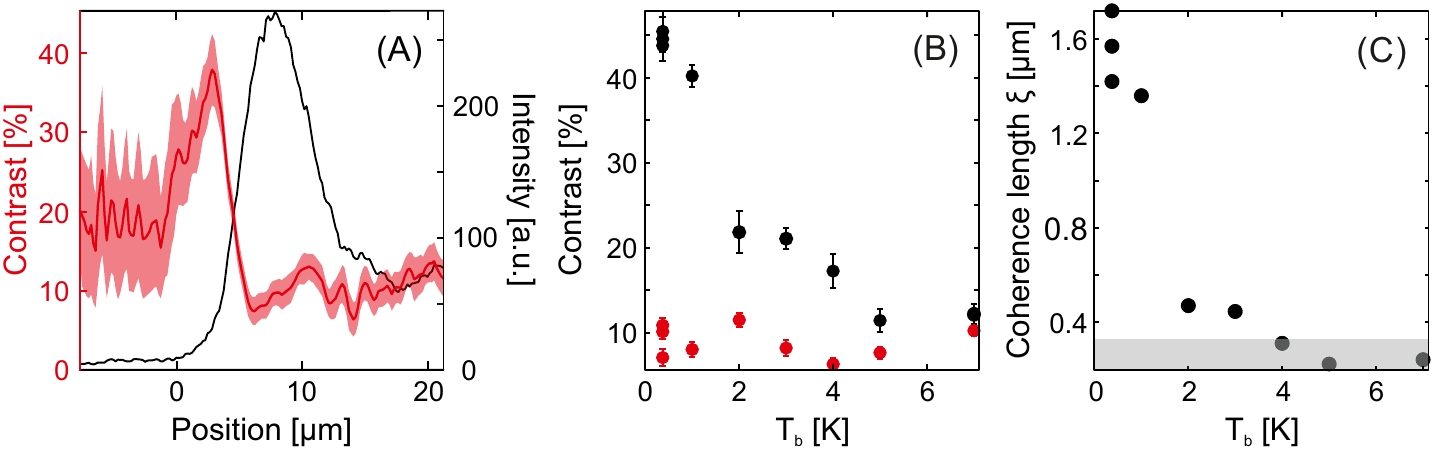}}}
\noindent {\bf Fig. 3.} (A): Intensity (dark) and interference contrast (red) of
the photoemission across the fragmented ring at 370 mK. The red area shows the
relative error in our measurements. The two vertical straight lines underline
the maximum of the emission, i.e., the position of the ring, and the maximum of
the interference contrast where the intensity is reduced by
10-fold. (B): Interference contrast for
$\delta_x$=1.5 $\mu$m as a function of the
bath temperature T$_\mathrm{b}$, taken 5 $\mu$m outside the ring (black) and at
the position of the ring (red). (C) Coherence length as a function of the bath 
temperature measured 5 $\mu$m outside of the ring. The gray area shows the limit of our experimental resolution.

\clearpage

\centerline{\scalebox{1.15}{\includegraphics[width=.9\textwidth]{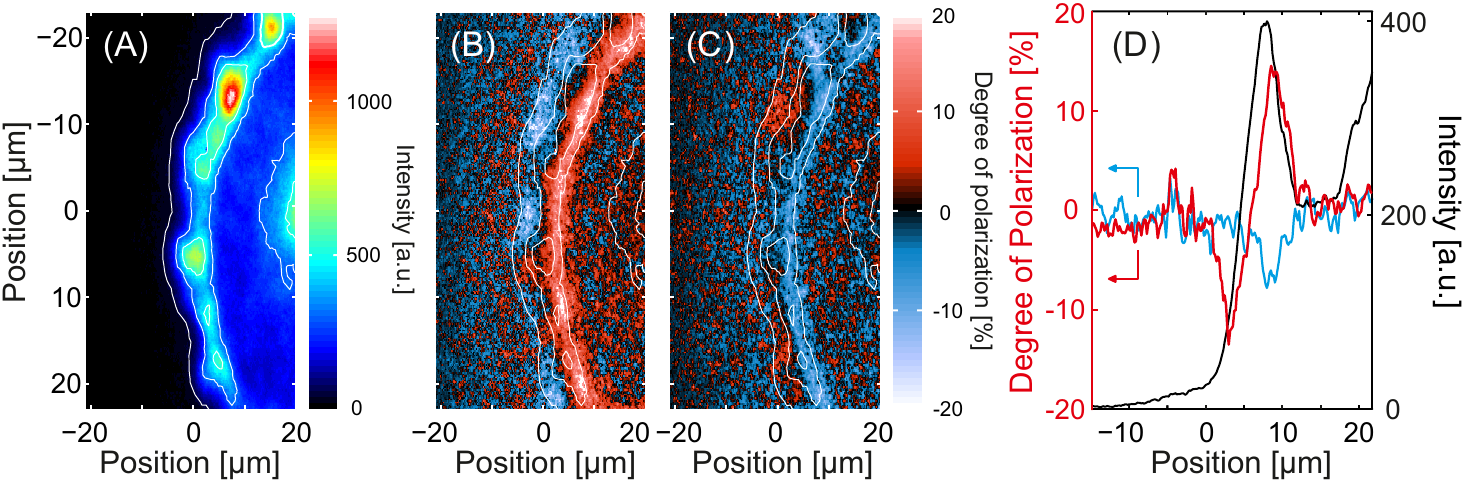}}}
\noindent {\bf Fig. 4.} Exciton ring at 370 mK (A). (B-C):
Pattern of linear (B) and circular (C) polarizations at 370 mK. In (C), the
blue and red colors are associated to $\sigma^+$ and $\sigma^-$ polarized
light respectively. In (B), we see that the photoemission is mostly linearly
polarized in the outer region of the ring where extended spatial coherence is observed. 
(D): Degree of linear (red) and circular (blue) polarization of the photoluminescence at 370 mK. The solid dark line shows the intensity profile of the emission. 

\newpage

\centerline{\large{\textbf{Supplementary Informations}}}

\textbf{Experimental details and sample structure}

Our sample is schematically shown in Figure S1. It consists of a 1 $\mu$m thick
field-effect device with an embedded 250
$\mathrm{\AA}$ wide
GaAs quantum well. The quantum well is surrounded
by Al$_{0.3}$Ga$_{0.7}$As barriers incorporating a ten-period (2/2 nm)
GaAs/Al$_{0.3}$Ga$_{0.7}$As superlattice.
The quantum well is placed 900 nm below a millimeter wide
semi-transparent gate electrode deposited on the surface of the sample. In our
experiments, 
this electrode is biased at a constant voltage
V$_\mathrm{g}\sim$-4.7 V with respect to the $n$-doped sample substrate that is
grounded. Our field-effect device is mounted in a
$^4$He-$^3$He cryostat, which limits the optical resolution
of our experiments to $\approx$ 1.5 $\mu$m due to mechanical vibrations. Let us
note that this value refers to the measured
full-width at half maximum of our gaussian-shaped point spread function.

Figure S2.B presents typical photoluminescence spectra emitted at the position
of the ring at low (350 mK) and high (7K) bath temperatures. One notes that
the emission spectrum narrows and acquires a well defined profile
as the bath temperature is lowered. Precisely, at 350 mK it is made of two
narrowband lines at an energy depending linearly on the
gate voltage applied on our top electrode (see Figure S2.C). This 
reveals that these two emissions result from the 
recombination of quasi-particles with an electric dipole moment $\sim$ 15(3) nm,
i.e. as expected for dipolar excitons confined in our electrically biased
quantum well
of 25 nm width. More interestingly, we also show that the
intensity ratio between these two emissions depends strongly on
the position on our sample: Figure S2.A shows that the high energy
line dominates the spectrum in the vicinity of the ring where charge neutrality is
best fulfilled. On the other hand, the ratio is essentially even in the inner region of
the ring which is rich in holes. This leads us to attribute 
the high energy component of the spectrum to the radiative recombination
of dipolar excitons while the line at lower energy is attributed to the interaction
between neutral excitons and excess carriers, e.g. holes in the inner region of
the ring (a more quantitative analysis of the emission spectrum is beyond the
scope of the present work and will be actually discussed elsewhere). Finally, we
also display in Fig. S2.A the photoluminescence spectrum emitted at the
position of the ring and 6 $\mu$m outside. These two emission profiles are essentially
identical, e.g. the spectral linewidth is unchanged between these two positions whereas the photoluminescence intensity varies by over one order of magnitude. 
This observation further confirms our conclusion of a weakly varying exciton
density in the electrostatic trap formed spontaneously in the vicinity of the
ring.  Otherwise the photoluminescence spectral width would vary as the result of homogeneous broadening which reflects the density of dipolar excitons\cite{Voros}.

\textbf{First order coherence of the photoluminescence}

To quantify the first order coherence of dipolar excitons, we magnify and
spatially filter the photoluminescence emitted in the ring region.
This part of the emission is directed towards a 
Mach-Zehnder interferometer with a path length difference
between its two arms that is actively stabilized.
This allows us to control the phase $\phi$ of the output interference signal
with a $\approx(\pi/10)$ accuracy. The magnified photoluminescence is then
split between the arms 1 and 2 of the interferometer, and a
vertical tilt angle ($\alpha$) is introduced between the outputs of the two
arms. Hence,
interference
fringes are aligned horizontally\cite{High_2012,Alloing_2012}, $\alpha$ being
set such that
the interference period is $\approx$ 4 $\mu$m. In this situation,
the outputs produced by the two arms are laterally shifted by $\delta_x$
while the path length difference remains close to zero. This allows us to derive
the 
degree of spatial coherence of the bright excitons. 

The output of our
interferometer, I$_{12}$, can be modelled as 
\newline
\begin{equation}
  I_{12}(\textbf{r};\delta_x)=\langle|\psi_0(\textbf{r},t)+e^{i(q_\alpha
y+\phi)}\psi_0(\textbf{r}+\delta_x,t)|^2\rangle_t,\nonumber
\end{equation}
where $\psi_0$(\textbf{r}) is the photoluminescence field which reflects the
bright excitons wave function, $\langle..\rangle_t$ denotes the time averaging
, \textbf{r}=(x,y) is the coordinate in the plane of
the quantum well
while
$q_\alpha$=2$\pi\lambda^{-1}$sin$(\alpha)$
where $\lambda$ is the emission wavelength. By recording individually the
output of the two arms, I$_{1}$ and I$_{2}$, we can compute interferograms or 
normalized interference patterns 
I$_\mathrm{int}$=(I$_{12}$-I$_{1}$-I$_2$)/2$\sqrt{I_{1}I_2}$ which
reveal
the first order coherence function of indirect excitons, defined as
\begin{equation}
 g^{(1)}(\textbf{r};\delta
x)=\frac{\langle\psi^*_0(\textbf{r},t)\psi_0(\textbf{r}+\delta_x,t)\rangle_t}{
(\langle|\psi_0(\textbf{r}
,t)|^2\rangle_t\langle|\psi_0(\textbf { r }
+\delta_x,t)|^2\rangle_t)^{1/2}}.\nonumber
\end{equation}
Indeed, I$_\mathrm{int}$(\textbf{r};$\delta_x$)=cos($q_\alpha
y+\phi+\phi_\mathrm{\textbf{r}}$)$|g^{(1)}(\textbf{r};\delta
x)|$ where
$\phi_\mathrm{\textbf{r}}$ is the phase of $g^{(1)}$. 
Hence, normalized interferences have a visibility controlled by the degree 
of spatial coherence of bright excitons, while the position of the
interference
fringes reveals the phase of $g^{(1)}$, i.e., the phase difference
between the interfering excitonic wave functions.

In Figure S3, we present some interferograms used to study the spatial
coherence of dipolar excitons (Figure 2). Varying the phase of the interferomer,
$\phi$, we deduce $|g^{(1)}(\textbf{r};\delta_x)|$ as shown in Fig. 2.C-D.

Our experiments show regions where the intensity of the photoluminescence
is weak while the degree of spatial coherence is large. To quantify it, we use a
standard numerical routine to fit the variation of I$_\mathrm{int}$ as a
function of $\phi$. The fitting error
becomes large when the intensity of the photoluminescence emission reaches
1$\%$ of its maximum at the ring position. This corresponds to $\approx$
10-20
counts on our photo-detector. Such a limitation is illustrated in Fig. S4
which
presents the map of the interference contrast together with its relative error
for the
experiment shown in Fig. 2.C. We see that, in the outer region of the ring,
the interference contrast is large and measured with a good precision 
while, for lower intensities, the fitting error is of the same order as the
contrast
we extract. To produce "simple" maps of the photoluminescence spatial
coherence (Figure 2.C-D), we discard the regions where the fitting error is
large. 

Finally, to extract the bright exciton coherence length $\xi$  quantitatively,
we model the variation of $|g^{(1)}|$ by the convolution between a Gaussian
profile
with a 1.5 $\mu$m full-width at half-maximum and an exponential
decay\cite{Fogler_2008,Semkat_2012}. 
The former function accounts for
our instrumental resolution while the latter function provides the theoretical
variation of $g^{(1)}$, namely $|g^{(1)}(\delta_x)|\propto e^{-\delta
_x/\xi}$. 
For our interferometric setup, we calibrate the variation of $|g^{(1)}|$ as a
function $\delta_x$ 
in previous experiments\cite{Alloing_2012}. The coherence
length of bright excitons $\xi$ is then obtained from the interference
contrast at $\delta_x$=1.5 $\mu$m (Fig. 2.H).

\textbf{Limitation of shift interferometry}

Semkat et al. have recently proposed theoretically that a sharp enough
photoluminescence pattern can lead to a significant visibility in shift-interferometry, even for a
spatially incoherent emission source~\cite{Semkat_2012}. To discard such effects
in our experiments,  we simulate the ring photoluminescence (which is the
sharpest pattern of the emission we study) with the incoherent photoluminescence
emitted by the n-doped GaAs substrate of our field-effect device. For that, we
excite the sample with a linear laser beam whose horizontal spatial linewidth
is varied. As predicted by Semkat et al., we verify that the interference
contrast is enhanced when the spatial width of the laser excitation is decreased.
Precisely, we obtain a contrast up to $\sim$15\% for a photoluminescence with
spatial extension $\sim 3.5\mu$m. Yet the visibility decreases rapidly when the
photoluminescence broadens spatially. This is shown in Fig.S5 for a
spatial extension $\sim 7.4\mu$m which is slightly smaller than the spatial
width of the ring at our lowest bath temperature. In this range, the
interference contrast is everywhere less than 10\% which sets the limit of our
instrumental resolution, in good agreement with the interference contrasts we
measure at high temperatures (see Fig. 3). Accordingly, we conclude that the
$\sim$ 45 $\%$ interference visibility we observe at sub-Kelvin bath
temperatures can not be explained by our instrumental resolution.



\clearpage

\centerline{\scalebox{1}{\includegraphics{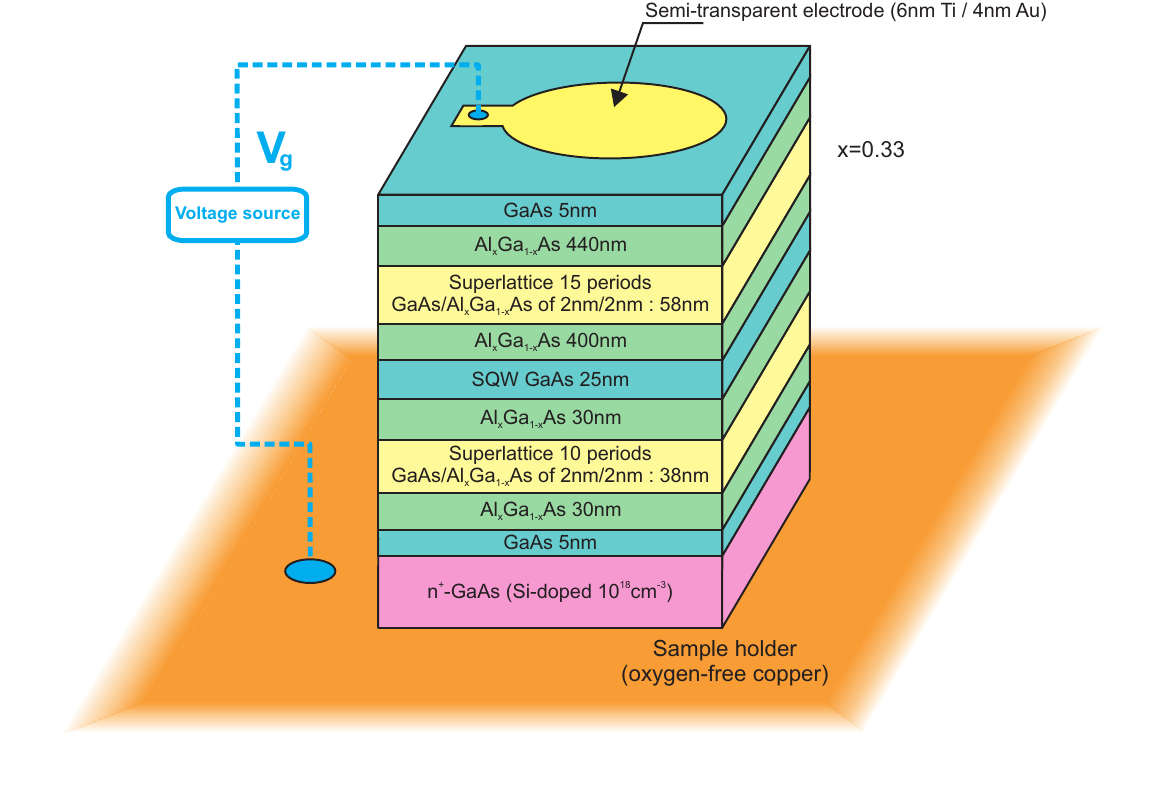}}}
\noindent {\bf Fig. S1.} 
\normalfont  Schematic representation of the field-effect device probed in
our experiments. 

\clearpage

\centerline{\scalebox{1}{\includegraphics{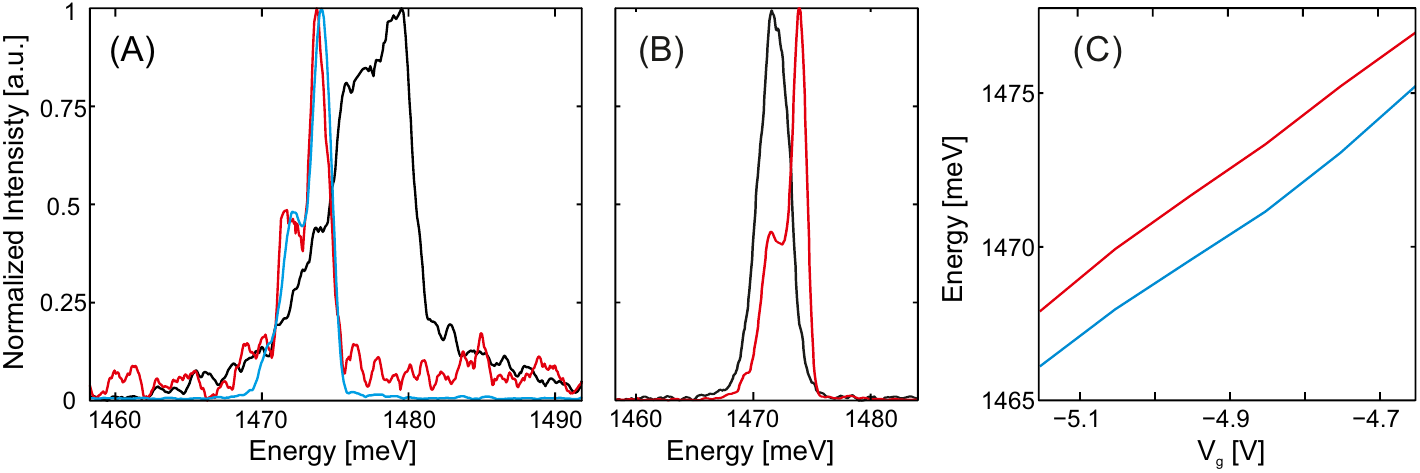}}}
\noindent {\bf Fig. S2.} 
\normalfont  (A): Normalized photoluminescence spectrum emitted at the position
of the ring (blue), in the inner region of the ring (black) and 6 $\mu$m outside the ring where we identify the "gray" condensate (red). Beside a factor 20 in intensity this spectrum is essentially identical to the one emitted at the position of the ring. Data are all acquired at 350 mK. 
(B): Rescaled photoluminescence spectra emitted at the position of the
ring at 350 mK (red) and at 7K (black). (C): Energy of the photoluminescence emitted at the position of the ring and as a function of the potential V$_g$ applied onto the top gate of our field-effect device. The variation of the low and high energy components of the spectrum are displayed in blue and red respectively. From these linear variations we deduce an electrical dipole moment $\sim$ 15(3) nm for both lines
at 350 mK. 

\clearpage

\centerline{\scalebox{1.2}{\includegraphics{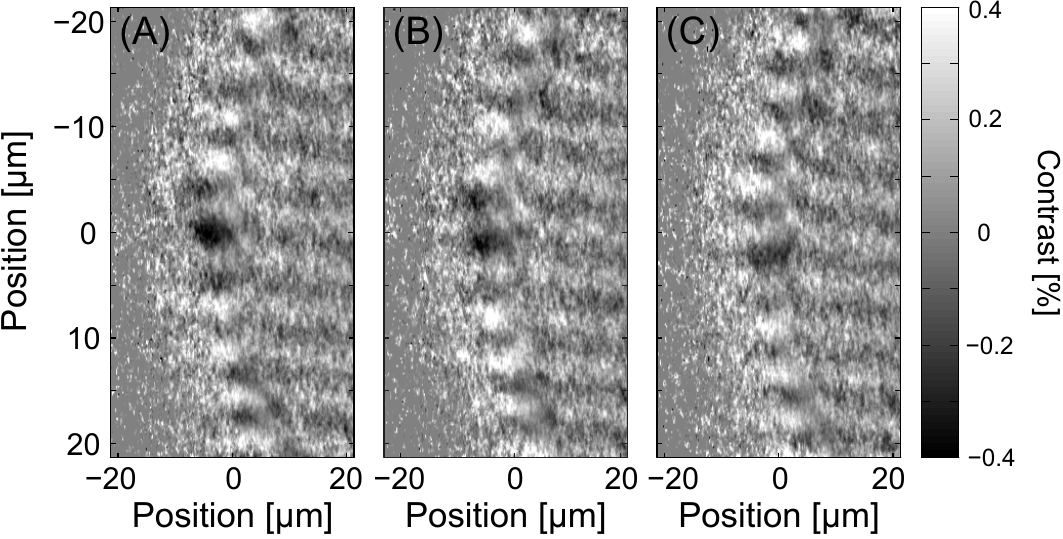}}}
\noindent {\bf Fig. S3.} 
\normalfont  (A-C): Normalized interference patterns $I_\mathrm{int}$ obtained
for $\delta_x$= 1.5 $\mu$m. From (A) to (C), the phase of the interferometer
$\phi$ are 0, $\pi$/2, 
and $\pi$. From such interferograms, we compute the map of $|g^{(1)}|$ shown in
Fig. 2.C. All measurements are realized at 370 mK by integrating in a 40 ns long time
interval set 10 ns after extinction of a
1 $\mu$s laser excitation.

\clearpage
\begin{center}
\includegraphics[width=1\textwidth]{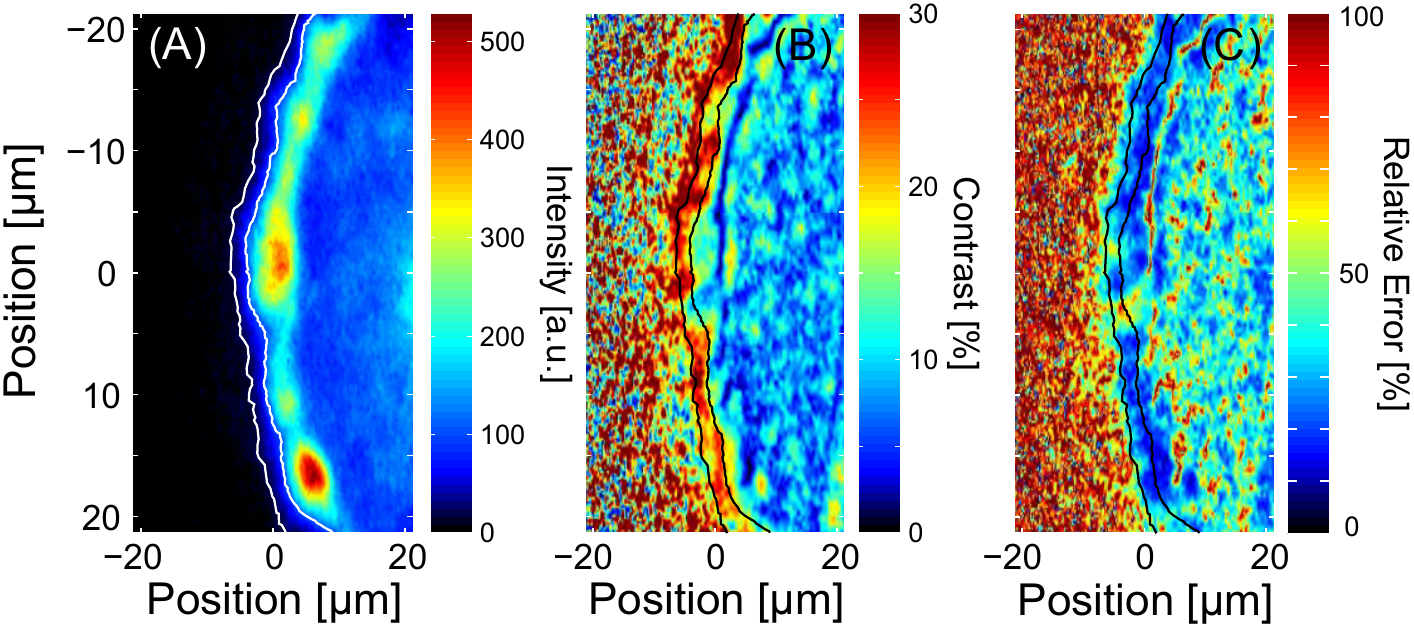}
\end{center}
\noindent {\bf Fig. S4.} 
\normalfont (A): Photoluminescence emission in the region of the fragmented ring
at 370 mK. This image is
identical to the one shown in Fig.2.A. (B): Map of the interference contrast
that we obtain by fitting point by point
the variation of the interference signal as a function of $\phi$. The relative
error of the fit is displayed in (C)
which shows that we can not extract the contrast with a sufficient accuracy for
regions where the photoluminescence 
intensity is less than $\approx$ 20 counts. In (A-C), 
the two contour lines are iso-intensities where the emission is
equal to 20 and 100 counts respectively. These underline the region where
we observe macroscopic spatial coherence with high resolution.

\clearpage
\begin{center}
\includegraphics[width=1\textwidth]{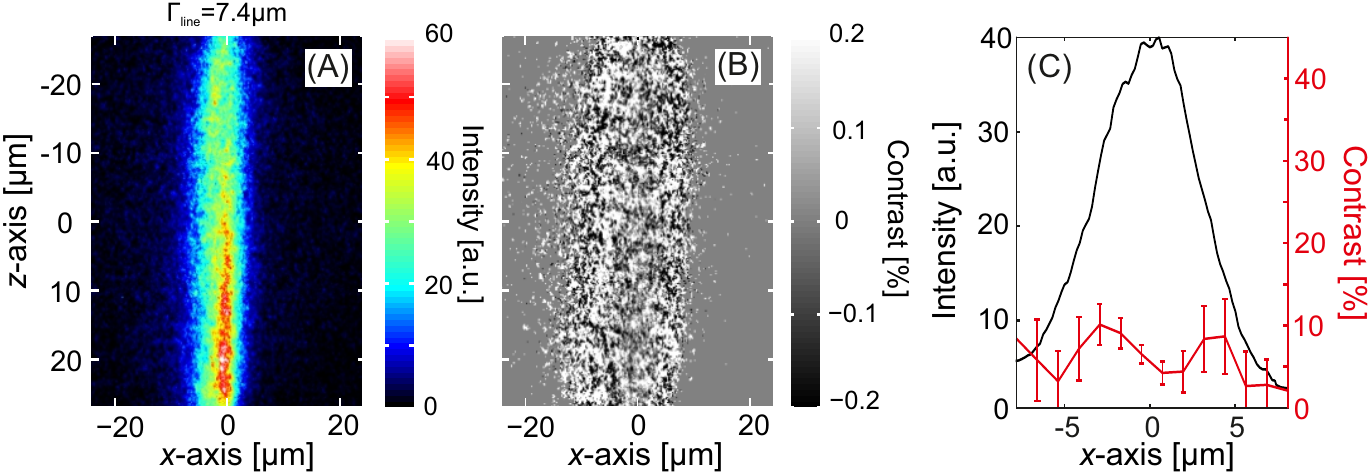}
\end{center}
\noindent {\bf Fig. S5.} 
\normalfont (A): Photoluminescence emission of the n-doped GaAs substrate
(V$_{\mathrm{g}}=0$) excited with a linear laser beam. The spatial width of the
photoluminescence along the $x$-axis is 7.4$\mu$m. (B) Normalized interferogram
obtained with this incoherent photoluminescence pattern for a lateral shift of
$\delta_{x}=1.5\mu$m. (C) Intensity profile of (A) (black line) and spatially
resolved contrast (red line), both along the $x$-axis. The contrast was obtained
by fitting $z$-axis profiles of the interferogram with a cosine function for
various positions on the $x$-axis. The intensity and contrast profiles are
limited to $\sim 10\mu$m on both sides of the origin of the horizontal axis,
for larger distances, the intensity of the photoluminescence is too weak
to resolve clear interferences. The ordinate scale of the interference contrast
in (C) is
set up to 45\% in order to match Fig. 3 of the main text.


\newpage

\begin{thebibliography}{99}

\bibitem{Stringari_Book_BEC} \textit{``Bose-Einstein
Condensation''}, L.P. Pitaevskii, S. Stringari (Oxford university Press,
2003)

\bibitem{Moskalenko_62} S.A. Moskalenko, Fiz. Tverd. Tela (Leningrad)
\textbf{4}, 276 (1962)

\bibitem{Blatt_1962} J. M. Blatt et al., Phys. Rev. \textbf{126}, 1691 (1962)

\bibitem{Keldysh_BCS} L.V. Keldysh and YuV. Kopaev, Sov. Phys. Solid State
\textbf{6}, 2219 (1965)

\bibitem{Keldysh_BEC} L.V. Keldysh and A.N. Kozlov, Sov. Phys. JETP \textbf{27},
521 (1968)

\bibitem{Stringari_Book} \textit{``Bose-Einstein condensation''}, Eds.
A. Griffin, D. W. Snoke, S. Stringari (Cambridge Univ. Press, 1995)

\bibitem{Snoke_Book} \textit{``Bose-Einstein
Condensation of Excitons and Biexcitons''}, S.A. Moskalenko, D. W. Snoke, 
(Cambridge Univ. Press, 2000)

\bibitem{Monique_dark_BEC} M. Combescot, O. Betbeder-Matibet, R. Combescot,
Phys. Rev. Lett. \textbf{99}, 176403 (2007)

\bibitem{Roland_2012} R. Combescot, M. Combescot, Phys. Rev. Lett. \textbf{109},
026401 (2012)

\bibitem{Monique_Leuenberger} M. Combescot, M.N. Leuenberger, Solid State Comm.
\textbf{149}, 567 (2009)

\bibitem{Deng_2002} H. Deng et al., Science \textbf{298}, 199 (2002)

\bibitem{Kasprzak_2006} J. Kasprzak et al., Nature \textbf{443}, 409 (2006)

\bibitem{Balili_2007} R. Balili et al., Science \textbf{316}, 1007 (2007) 

\bibitem{Wertz_2010} E. Wertz et al., Nat. Phys. \textbf{6}, 860–864 (2010)

\bibitem{Leggett_2006} A.J. Leggett \textit{Quantum Liquids} (Oxford. Univ.
Press, 2006)  

\bibitem{Ueda_2013} D.M. Stamper-Kurn, M. Ueda, Rev. Mod. Phys. \textbf{85},
1191 (2013) 

\bibitem{Supplements} see Supplementary Informations

\bibitem{Blackwood_94} E. Blackwood et al., Phys. Rev. B \textbf{50}, 14246
(1994) 

\bibitem{Timofeev_2013}	A.V. Gorbunov, V.P. Timofeev, Sol. Stat. Comm.
\textbf{157}, 6 (2013)

\bibitem{Butov_2004} L. V. Butov et al., Phys. Rev. Lett \textbf{92}, 117404
(2004)

\bibitem{Rapaport_2004} R. Rapaport et al., Phys. Rev. Lett \textbf{92}, 117405
(2004)

\bibitem{Butov_2002} L.V. Butov et al., Nature \textbf{418}, 751 (2002)

\bibitem{Snoke_2002} D.W. Snoke et al., Nature \textbf{418}, 754 (2002)

\bibitem{Alloing_Sci_rep} M. Alloing, A. Lemaitre, E. Gallopin, F. Dubin, Sci. Rep. \textbf{3}, 1578 (2013)

\bibitem{Ivanov_2010} A. L. Ivanov, E. A. Muljarov, L. Mouchliadis, and
R. Zimmermann, Phys. Rev. Lett. \textbf{104}, 179701 (2010)

\bibitem{Schindler_08} C. Schindler and R. Zimmermann, Phys. Rev. B \textbf{78},
045313 (2008)

\bibitem{Rapaport_09} B. Laikhtman. and R. Rapaport, Phys. Rev. B \textbf{80},
195313 (2009)

\bibitem{Glauber_1999} M. Naraschewski and R. J. Glauber, Phys. Rev. A \textbf{59}, 4595 (1999)

\bibitem{IB_2000} I. Bloch, T. W. Haensch,  T. Esslinger, Nature \textbf{403}, 166 (2000)


\bibitem{High_2012} A.A. High et al., Nature \textbf{483}, 584 (2012)

\bibitem{Ali_2009} M. Ali Can and T. Hakioglu, Phys. Rev. Lett. \textbf{103},
086404 (2009)

\bibitem{Kavokin_2012} M. Matuszewski et al., Phys. Rev. B \textbf{86}, 115321
(2012)

\bibitem{Malpuech_2013} D.V. Vishnevsky et al., Phys. Rev. Lett. \textbf{110},
246404 (2013)

\bibitem{High_2013} A.A. High, et al., Phys. Rev. Lett. \textbf{110}, 246403
(2013)




\end{thebibliography}

\begin{thebibliography}{99}


\bibitem{Voros}  Z. Voros, D.W. Snoke, K. West and L. Pfeiffer, Phys. Rev. Lett.
\textbf{103}, 016403 (2009)


\bibitem{High_2012} A.A. High et al., Nature \textbf{483}, 584 (2012)

\bibitem{Alloing_2012} M. Alloing, D. Fuster, Y. Gonzalez, L. Gonzalez and F.
Dubin, arXiv:1210.3176

\bibitem{Fogler_2008} M.M. Fogler et al.,  Phys. Rev. B \textbf{78}, 035411
(2008)

\bibitem{Semkat_2012} D. Semkat et al.,  Nano Lett. \textbf{12}, 5055 (2012)



\end{thebibliography}
\end{document}